\documentclass[11pt]{amsart}
\usepackage{fullpage}
\usepackage{amsthm,amsmath,amssymb}
\usepackage[foot]{amsaddr}

\usepackage{enumitem}
\usepackage{booktabs}
\usepackage{graphicx}
\usepackage{subcaption}
\usepackage{tikz}
\usepackage{natbib}

\usepackage[colorlinks=true,linkcolor=red,citecolor=blue,urlcolor=cyan]{hyperref}

\graphicspath{{./figs/}}

\newtheorem{theorem}{Theorem}[section]
\newtheorem{corollary}[theorem]{Corollary}
\newtheorem{proposition}[theorem]{Proposition}
\newtheorem{lemma}[theorem]{Lemma}
\theoremstyle{remark}
\newtheorem{rem}{Remark}

\newcommand{\ie}{{\it i.e.}}
\newcommand{\eg}{{\it e.g.}}
\newcommand{\cf}{{\it c.f.~}}
\DeclareMathOperator{\sgn}{sgn}

\title{Quantifying Sufficient Randomness \\ for Causal Inference}

\author{Brian Knaeble}
\address{Department of Computer Science, Utah Valley University, Orem, UT}
\email{bknaeble@uvu.edu}

\author{Braxton Osting}
\address{Department of Mathematics, University of Utah, Salt Lake City, UT}
\email{osting@math.utah.edu}

\author{Placede Tshiaba}
\address{Department of Mathematics, University of Utah, Salt Lake City, UT}
\email{placede.tshiaba@utah.edu}


\begin{document}
\maketitle

\begin{abstract}
Spurious association arises from covariance between propensity for the treatment and individual risk for the outcome. For sensitivity analysis with stochastic counterfactuals we introduce a methodology to characterize uncertainty in causal inference from natural experiments and quasi-experiments. Our sensitivity parameters are standardized measures of variation in propensity and individual risk, and one minus their geometric mean is an intuitive measure of randomness in the data generating process. Within our latent propensity-risk model, we show how to compute from contingency table data a threshold, $T$, of sufficient randomness for causal inference. If the actual randomness of the data generating process exceeds this threshold then causal inference is warranted.
\end{abstract}

\section{Introduction}
Causal inference from observational data is a challenging problem. Even an apparently homogeneous population will contain individuals with unmeasured yet distinguishing characteristics, and these characteristics can lead to spurious association. One approach to this problem of unmeasured confounding is to select and condition on conveniently measured covariates. However, it is not clear when a particular covariate should be included in the model \citep{DM} and, of course, there are always additional unmeasured covariates to consider. 
To approach these challenges, statisticians have introduced the ideas of propensity for treatment and sensitivity analysis. 
A \emph{propensity} is the individual probability of treatment, which is estimated from population data with a \emph{propensity score}. 
A \emph{sensitivity analysis} attempts to quantify how an inference depends on statistical modeling assumptions, often emphasizing omitted variables bias. 
A practitioner may repeatedly fit different candidate models to assess the sensitivity of an estimate, while a theoretician will identify properties of unmeasured variables that are both intuitive and informative of sensitivity. An approach that is both theoretically sophisticated and applicable in practice is to design observational studies to mimic randomized experiments, in part because randomization tends to balance both measured and \emph{unmeasured} covariate data \citep[p. 73]{Rose10}. 

\emph{The goal of this paper is to quantify uncertainty in causal inference from natural experiments and quasi-experiments.} Instead of asking how exposed and unexposed patients differ with  regards  to  measured and  unmeasured covariate data,  we  ask \emph{how stochastic, \ie, non-deterministic, are the processes assigning treatment and the outcome?} To address this question,  we introduce a \emph{latent propensity-risk model} of the data generating process. Similar to previous methods, propensity is the propensity for treatment (or exposure). But, since outcome is also a chance event, we include a second propensity for the outcome (or disease), which we refer to as \emph{individual risk}. We thus have a symmetric framework for sensitivity analysis with stochastic counterfactuals, \cf \citet{VR}. Our model emphasizes the population distribution of propensity-risk and de-emphasisizes covariate data. This model is formulated in Section~\ref{methods}. 

In our latent propensity-risk model, covariance between propensity and risk is a reparametrization of spurious association between treatment and outcome. Thus, we reason that when there is hidden covariance between propensity and risk, spurious association arises in the observable data. Conversely, given an observed association we can ask how much covariance is needed to explain it away. Two intuitive factors of latent covariance are square roots of separate coefficients of determination, and a measure of stochasticity is one minus their geometric mean, which we refer to as the \emph{randomness} of the data generating process. A constrained optimization problem over the space of propensity-risk distributions asks for the maximum randomness consistent with the observed data under a null hypothesis that the treatment does not cause the outcome; see Equation \eqref{e:OptProblem}. We refer to the maximum randomness value, consistent with the observed data, as a \emph{threshold}, $T$, of sufficient randomness for causal inference. In Section~\ref{results}, we show how to compute the sensitivity parameter $T$ from observed contingency table data. In particular, we show that $T$ can be determined from a measure of association known as the $\phi$ coefficient; see  Theorem~\ref{abcd}. If our propensity-risk model is appropriate and it can be demonstrated that the true randomness of the data generating process exceeds the computed $T$ value then causal inference is warranted. Interestingly, full randomization is not required. The sensitivity parameter $T$ depends on measured association, and also prevalence of treatment and prevalence of outcome. We provide formulas for computing $T$ from measurements of prevalence and commonly used measures of association including the risk difference ($RD$), the relative risk ($RR$), and the odds ratio ($OR$); see Proposition~\ref{P}.

In Section~\ref{applications}, we give two example applications illustrating how to compute the threshold $T$ of sufficient randomness from contingency table data. We demonstrate how it can support accurate causal interpretation of  an  observed  association.   In  the  first  application,  we  see  that  a  large  amount  of  randomness is needed,  even though the association is large.  In the second application we see that a smaller amount of randomness is needed, even though the association is smaller. Each measure of association $RD$, $RR$, and $OR$ by itself is an inadequate gauge of sensitivity. For the purpose of assessing sensitivity the measure of association $\phi$ appears to be more directly applicable.

In Section~\ref{discussion} we further discuss our introduced methodology and future directions.

\subsection{Comparison to previous work} 
Sensitivity analysis with our randomness threshold $T$ is most similar to the sensitivity analysis of \citet[Chapter 4]{Rose}. 
For pairs of apparently identical individuals \citet{Rose10} considers the odds ratio of their treatment propensities, and his sensitivity parameter $\Gamma$ is the least upper bound of these odds ratios. Our proposed methodology is complementary, and in Section~\ref{discussion} we discuss extensions of our framework to handle measured covariate data and application when sample sizes are small. Our emphasis on randomness is consistent with Fisher's view of randomization as the ``reasoned basis'' for causal inference \citep[p. 37]{Fisher,Rose10}.

Our method is based on the methodologies of \citet{KD} and \citet{KOA} for sensitivity analysis of continuous data. Related methodologies for sensitivity analysis of continuous data with coefficients of determination as sensitivity parameters are described in \citet{HHH}, \citet{Frank}, \citet{CH}, and \citet{Oster}. The seed of our methodology was the mathematical symmetry in the analysis of \citet{KOA}. Based on a series of interviews with practicing epidemiologists, the methodologies of \citet{KD} and \citet{KOA} have been adapted and modified for practical application during analysis of categorical data, resulting in the methodology of this paper. Our main formula in Section~\ref{results} can be applied directly to observed relative frequencies of a contingency table. 

The methodology of \citet{KD} deals with two coefficients of determination, while the methodology of \citet{SAWA} deals with two relative risks (risk ratios). \cite{EV} introduce the E-value as the minimum strength of association on the risk ratio scale that an unmeasured confounder would need to have with both the treatment and the outcome to fully explain away a specific treatment-outcome association,
conditional on the measured covariates. While the E-value is a property of omitted covariates, the randomness is a property of the data generating process.

\subsection{Outline} 
In Section~\ref{methods}, we formulate the 
latent propensity-risk model and our sensitivity threshold $T$. 
In Section~\ref{results}, we show how the sensitivity threshold $T$ of sufficient randomness can be computed from observed contingency table data. 
In Section~\ref{applications}, we discuss two example applications illustrating how to compute the threshold $T$. We conclude in Section~\ref{discussion} with a discussion. 
Proofs of the mathematical results given in this paper are given in Appendix~\ref{s:Proofs}.

\section{Methods}
\label{methods}
We assume an infinite population of independent individuals, each with a propensity $p$ for an exposure $e$, and a potential risk $r$ for a disease $d$ in the counterfactual absence of $e$. For each individual, in the absence of $e$ the risk is $r$, and in the presence of $e$ the risk is $r+\beta$, where $\beta$ is their individual causal effect. 
We are not \emph{assuming} an additive effect, nor are we assuming a constant causal effect. The random variable $\beta$ simply represents for each individual the difference between their risk in the presence of exposure and their risk in the absence of exposure.
We think of $r$ as risk for $d$ at a time prior to when $e$ could occur. Over the whole population we say that $e$ causes $d$ if $\beta\neq 0$ on a nonzero proportion of individuals. Our null hypothesis is that $e$ does not cause $d$, \ie,
\begin{equation*}
H_0: P(\beta=0)=1.
\end{equation*} 
It is understood that the exposure $e$ is an indicator for any general event thought to be causal, \eg, a treatment, and $d$ measures some general dichotomous categorical response, \eg, an outcome.

\subsection{The randomness of the data generating process}
\label{Rdefshere}
Under $H_0$ our propensity-risk model of the data generating process is described as follows: for each individual, both measured and unmeasured covariates determine $(p,r)$, and then chance assigns $(e,d)$. Across a population of individuals we have a distribution of $(p,r)$ values. Let $\mathcal{P}(S)$ be the space of $(p,r)$ distributions on the open unit square, $S= (0,1)^2$. We write $\mu$ for a generic distribution in $\mathcal{P}(S)$. The randomness of the data generating process is formally defined from the first and second marginal moments of $\mu$. With $E = E_\mu$ for expectation with respect to the distribution $\mu$, and $\sigma^2$ for variance of the subscripted variable, we may compute 
$E[p] =\int_S p  \ d\mu$, 
$E[r]=\int_S r  \ d\mu$, 
$\sigma^2_p = \int_S (p-E[p])^2 \ d\mu$, and 
$\sigma^2_r= \int_S (r-E[r])^2 \ d\mu$.
We then define
\begin{equation*}
    \label{Rdef} 
    R^2_p=\frac{\sigma^2_p}{E[p](1-E[p])} \quad \textrm{and} \quad R^2_r=\frac{\sigma^2_r}{E[r](1-E[r])},
\end{equation*}
and their geometric mean \begin{equation*}\label{randomness1}R^2=(R^2_pR^2_r)^{1/2}=R_pR_r.\end{equation*} We refer to 
\begin{equation}
\label{randomness2}
\eta=1-R^2
\end{equation} 
as the \emph{randomness} of the data generating process. We have $R^2\in[0,1)$ and $\eta\in(0,1]$.

\begin{rem}
In defining the randomness, we use the geometric mean of $R_p^2$ and $R_r^2$. The arithmetic mean $\frac{R_p^2+R_r^2}{2}$ is insufficient for our purposes, while the geometric mean $R_pR_r$ is a sufficient summary of $(R^2_p,R^2_r)$; see Section~\ref{discussion}. We describe an alternate formulation of $R^2$ in \eqref{interesting1}.
\end{rem}

\begin{figure}[b!]
\centering 
\includegraphics[width=4.0in]{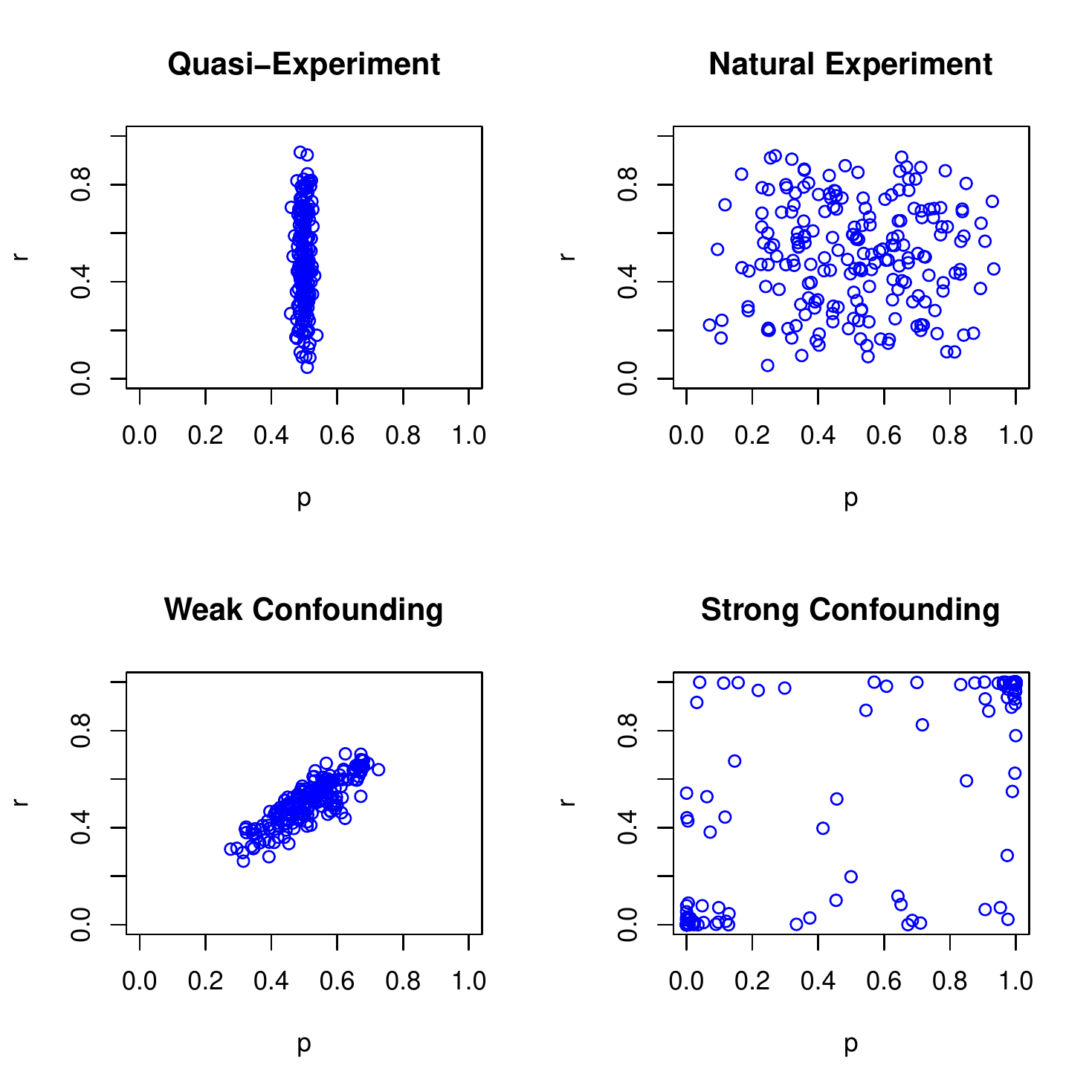}
\caption{We plot four examples of latent, propensity-risk distributions. 
These latent distributions are a thought experiment; we do not estimate the latent distributions from data. 
In the top two figures spurious association is absent because there is no correlation between latent propensity and risk. 
In the bottom two figures spurious association arises due to correlation between latent propensity and risk.
Strong confounding is possible when there is excess variation in propensity and risk, \ie, when $R^2$ is large (see Section~\ref{Rdefshere}). The randomness $\eta=1-R^2$ in the Weak Confounding example is $99\%$, while the randomness in the Strong Confounding example is only $13\%$. Measured and unmeasured covariates play a larger role when the distribution is dispersed and a smaller role when the distribution is concentrated near a point. Chance plays a larger role when the distribution is compressed, and in that case assignment to $e$ or $d$ is more ``random''. The randomness in the Quasi-Experiment example is $99\%$, and the randomness in the Natural-Experiment example is $83\%$.} 
\label{sampledistributions}
\end{figure}

The quantities $R^2_p$ and $R^2_r$ are generalized coefficients of determination. They are fractions of variation \emph{explainable} from measured and unmeasured attributes of individuals. The quantities $1-R^2_p$ and $1-R^2_r$ are fractions of variation \emph{unexplainable} from measured and unmeasured attributes, \ie, due to chance. We think of $R^2_p$ and $R^2_r$ as coefficients of \emph{determinism} and $1-R^2_p$ and $1-R^2_r$ as coefficients of randomness. These coefficients are intuitive enough to be specified from knowledge of the data generating process, and together they determine $\eta$ (see \eqref{randomness2}). We discuss specification of $R^2_p$ and $R^2_r$ in Section~\ref{discussion}. Here we note that $R^2_p$ is zero in a randomized experiment and near one in an observational study with highly deterministic assignment to $e$. $R^2_p$ (or $R^2_r$) is smaller when assignment to $e$ (or $d$) is more stochastic. Figure~\ref{sampledistributions} illustrates how knowledge of randomness in the data generating process can be used to bound the magnitude of spurious association.

\subsection{The threshold, $T$, of sufficient randomness}
\label{thresh}
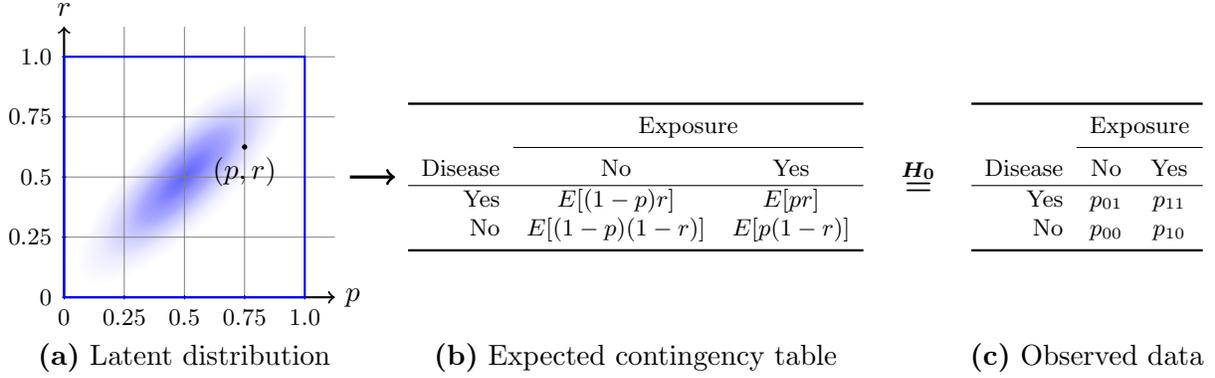
\begin{figure}[t!]
\begin{center}
\begin{tikzpicture}[scale=0.8]
\node[] at (2,-1){{\bf (a)} Latent distribution};
\foreach\i in {0,0.01,...,1} {\fill[opacity=\i*0.02,blue,rotate around={45:(2,2)t}] (2,2) ellipse ({3-3*\i} and {1-1*\i});   }
\draw[step=1cm,gray,very thin] (-0,-0) grid (4.5,4.5);
\draw[thick,->] (0,0) -- (4.5,0) node[anchor=west] {$p$};
\draw[thick,->] (0,0) -- (0,4.5) node[anchor=south] {$r$};
\foreach \x in {0,0.25,0.5,0.75,1.0}
   \draw (4*\x cm,1pt) -- (4*\x cm,-1pt) node[anchor=north] {\footnotesize $\x$};
\foreach \y in {0,0.25,0.5,0.75,1.0}
    \draw (1pt,4*\y cm) -- (-1pt,4*\y cm) node[anchor=east] {\footnotesize $\y$};
\draw[blue,thick] (0,0) -- (4,0) -- (4,4) -- (0,4) -- (0,0);
\filldraw[black] (3,2.5) circle (1pt) node[anchor= north ] {$(p,r)$};

\draw[very thick,->] (4.75,2) -- (5.5,2);

\node[] at (9.5,-1){{\bf (b)} Expected contingency table};
 \node[] (0) at (9.5,2) { \footnotesize
\begin{tabular}{rcc}
\toprule
 & \multicolumn{2}{c}{Exposure}\\
\cmidrule{2-3}
Disease&No&Yes\\
\hline
Yes& $E[(1-p)r]$ & $E[pr]$  \\
No& $E[(1-p)(1-r)]$ & $E[p(1-r)] $ \\
\bottomrule
\end{tabular}
  };

\node[] at (14.2,2.0) { \large $\stackrel{\boldsymbol{H_0}}{\boldsymbol{=}}$};

\node[] at (17,-1){{\bf (c)} Observed data};
 \node[] (0) at (17,2) { \footnotesize
\begin{tabular}{rcc}
\toprule
 & \multicolumn{2}{c}{Exposure}\\
\cmidrule{2-3}
Disease&No&Yes\\
\hline
Yes& $p_{01}$ & $p_{11}$  \\
No& $p_{00}$ & $p_{10}$ \\
\bottomrule
\end{tabular}
  };
\end{tikzpicture}
\caption{\label{f:PRmodel} 
{\bf (a) }An illustration of a latent propensity-risk distribution on the $(p,r)$ unit square and 
{\bf (b)} the resulting expected relative frequencies. Under $H_0$ these expected relative frequencies equal the observed relative frequencies shown in 
{\bf (c)}. The threshold, $T$, of sufficient randomness is computed under $H_0$ by finding the distribution with the largest randomness that is consistent with the observed data; see \eqref{SR}. If our propensity-risk model is accurate and the coefficients of determinism $(R^2_p,R^2_r)$ satisfy $1-R_pR_r>T$ then we reject $H_0$ and causal inference is warranted, \ie, we need to appeal to causality in order to explain the data.}
\end{center}
\end{figure}

From the observations of $(e,d)$ we know the relative frequencies
\[
p_{01}=P(e=0,d=1), \ \  
p_{11}=P(e=1,d=1), \ \  
p_{00}=P(e=0,d=0), \ \  \textrm{and} \ \ 
p_{10}=P(e=1,d=0).
\] 
Let $\mathcal P_0 (S) \subset \mathcal P (S)$ be the class of \emph{feasible} distributions satisfying \[E[(1-p)r]=p_{01},\quad E[pr]=p_{11},\quad E[(1-p)(1-r)]=p_{00}, \quad \textrm{and}\quad E[p(1-r)]=p_{10}.\]
$\mathcal P_0 (S)$ is the set of distributions that are consistent with the observed data (the observed contingency table) under $H_0$; see Figure~\ref{f:PRmodel}.

Based on \eqref{randomness2}, we define the \emph{threshold}, $T$, of sufficient randomness for causal inference by
\begin{equation}
\label{SR}
T := 1 - R^2_\star = 1-\min_{\mu\in \mathcal P_0 (S)}R^2(\mu)=\max_{\mu\in \mathcal P_0 (S)}\eta(\mu)=\eta^\star.
\end{equation}
$T$ is the maximum amount of randomness that is consistent with the observed data under $H_0$. If the coefficients of determinism $(R^2_p,R^2_r)$ satisfy $1-R_pR_r>T$ then causal inference is warranted.

\begin{rem} The threshold $T$ is computed from the observed data. Specifying $(R^2_p,R^2_r)$ is loosely analogous to specifying the size of a hypothesis test, where size refers to the probability of falsely rejecting a null hypothesis. Computation of $T$ is like computation of a p-value. We may reject $H_0$ if $T<1-R_pR_r$.
\end{rem}

To compute $T$ we solve the minimization problem in \eqref{SR}, which can be explicitly written as follows:
\begin{subequations}
\label{e:OptProblem}
\begin{align}
\label{e:OptProblema}
R^2_\star = \min_{\mu \in \mathcal{P}(S)} \ & R^2(\mu), 
\qquad \qquad 
\textrm{where } R^2(\mu) = \sqrt{\frac{\sigma^2_p(\mu) \ \sigma^2_r(\mu)} {E[p] \ \left(1-E[p]\right) \ E[r] \ \left( 1-E[r] \right) }}\\
\label{e:OptProblemb}
\text{such that} \  & E[(1-p)r]  \ = \ p_{01} \\
\label{e:OptProblemc}
& E[pr]   \ = \ p_{11} \\
\label{e:OptProblemd}
& E[(1-p)(1-r)]   \ = \ p_{00} \\
\label{e:OptProbleme}
& E[p(1-r)]   \ = \ p_{10} 
\end{align}
\end{subequations}
The objective function is the geometric mean of standardized variances, while the constraint set requires that $\mu$ be a feasible distribution on the $(p,r)$ unit square, \ie, the distribution $\mu$ is required to be consistent with the observed data under $H_0$.

\section{Results}
\label{results}
The following theorem gives an analytic solution to the optimization problem in \eqref{e:OptProblem} and hence an explicit formula for the threshold $T$ of sufficient randomness; see \eqref{SR}. 
\begin{theorem}
\label{abcd}
Suppose $P(e=1)\in (0,1)$ and $P(d=1)\in (0,1)$. Let $p_{01}=P(e=0,d=1)$, $p_{11}=P(e=1,d=1)$, $p_{00}=P(e=0,d=0)$, \textrm{and}\quad $p_{10}=P(e=1,d=0)$. Define 
\[\phi :=\frac{p_{11}p_{00} - p_{01}p_{10}}{\sqrt{(p_{11}+p_{10})(p_{00}+p_{10})(p_{01}+p_{11})(p_{01}+p_{00})}}.
\]
The threshold, $T$, of sufficient randomness for causal inference can be computed with the following formula: 
\begin{equation}
\label{tform}
T=1-|\phi|. 
\end{equation}
\end{theorem}
\noindent Theorem~\ref{abcd} is proven in Appendix~\ref{a1}. The quantity $\phi$ is a measure of association known as the {\it $\phi$ coefficient}. The $\phi$ coefficient is an analogue of Pearson's correlation coefficient, $\rho$, but $\phi$ measures association between two dichotomous, categorical variables. The coefficient $\phi$ is related to the chi-squared statistic, and we discuss this connection in Section~\ref{discussion}.

We proceed to describe formulas for computing $T$ from the observed prevalence of the exposure, the observed prevalence of the disease, and an observed measure of association. 
Write the standard (open) 3-simplex 
\begin{equation*}
\Delta := \left\{ (p_{01},p_{11},p_{00},p_{10})\in\mathbb{R}^4 \colon \ 
p_{01}, \ p_{11}, \ p_{00}, \ p_{10}>0; 
\ \  p_{01}+p_{11}+p_{00}+p_{10}=1 \right\}.
\end{equation*}
Denote the marginal relative frequencies by $p_e=P(e=1) = p_{11} + p_{10}$ and $p_d=P(d=1) = p_{11} + p_{01}$. 
The set $\Delta$ can be reparametrized with $(p_e,p_d,\alpha)$, where $\alpha$ is a measure of association. We have already considered $\alpha=\phi$. Next we consider common measures of association known as the risk difference (RD), 
\[RD=\frac{p_{11}}{(p_{11}+p_{10})}-\frac{p_{01}}{(p_{01}+p_{00})},\]
the relative risk (RR), 
\[RR=\frac{p_{11}}{(p_{11}+p_{10})}\frac{(p_{01}+p_{00})}{p_{01}},\]
and the odds ratio (OR),
\[OR=\frac{p_{11}}{p_{10}}\frac{p_{00}}{p_{01}}.\]

While Theorem~\ref{abcd} gives an explicit formula for $T$ in terms of the relative frequencies of a contingency table,  
the following proposition (Proposition~\ref{P}) provides formulas for computing $T$ from observed $(p_e,p_d,\alpha)$, where $\alpha\in\{RD,RR,OR\}$. 
We anticipate that it may be useful for gaining intuition for $T$ and  during meta analysis of summary statistics.

\begin{proposition}
\label{P}
Suppose $p_e,p_d\in(0,1)$. Define $k=\sqrt{\frac{p_e(1-p_e)}{p_d(1-p_d)}}$. For $\alpha \in \{RD,RR,OR\}$, 
\begin{equation}
\label{RDformula}
    T=
    \begin{cases}
 1-|RD|k & \textrm{if } \alpha=RD\\
  1- \left|\frac{p_d(RR-1)}{1+p_e(RR-1)} \right| k & \textrm{if } \alpha=RR\\
  1-\left|\frac{p_d(u-1)}{1+p_e(u-1)} \right| k & \textrm{if } \alpha=OR,\\
\end{cases}.
\end{equation}
where $u=\frac{-a+\sqrt{a^2-4 p_e c}}{2 p_e}$, $a = p_d\left(OR-1\right)+\left(1-p_e\right)-p_eOR$, and $c = (p_e-1)OR$.
\end{proposition}

\noindent Proposition~\ref{P} is proven in Appendix~\ref{a2}. 

\medskip
Next, we consider $T$ as a sensitivity parameter, and provide a qualitative summary of how $T$ depends on $(p_e,p_d,\alpha)$, where $\alpha\in\{RD,RR,OR\}$.  
We say that a triple $(p_e,p_d,\alpha)$ is \emph{realizable} if it could have resulted from relative frequencies $(p_{01},p_{11},p_{00},p_{10})\in \Delta$. Conditional on $\alpha$ we say that $(p_e,p_d)$ is realizable if $(p_e,p_d,\alpha)$ is realizable. Conditional on $(p_e,p_d)$ we say $\alpha$ is realizable if $(p_e,p_d,\alpha)$ is realizable. For $\alpha\in\{RD,RR,OR,\phi\}$ we write $l_\alpha$ for the greatest lower bound on realizable $\alpha$, and $u_{\alpha}$ for the least upper bound on realizable $\alpha$. 
The following lemma presents formulas for $l_\alpha$ and $u_\alpha$ for $\alpha\in\{RD,RR,OR\}$. 
\begin{lemma}\label{l:simp} 
The triple $(p_e,p_d,\alpha)$ is realizable if and only if $p_e,p_d\in(0,1)$ and  \newline
for $\alpha=RD$, 
\[-1\leq l_{RD}<RD< u_{RD}\leq1,\]
for $\alpha=RR$, 
\[0\leq l_{RR}<RR<u_{RR}\leq\infty, \textrm{~or}\]
for $\alpha=OR$,
\[0=l_{OR}<OR<u_{OR}=\infty,\]
where 
\begin{align*}
l_{RD} &= -\min\left\{\frac{p_d}{1-p_e},\frac{1-p_d}{p_e}\right\}, &&
u_{RD} =\min\left\{\frac{p_d}{p_e},\frac{1-p_d}{1-p_e}\right\}, \\
l_{RR} &=\max\left\{0,\frac{p_e+p_d-1}{p_e}\right\},  \quad \textrm{~and~} && 
u_{RR} =
\begin{cases}
\frac{1-p_e}{p_d-p_e} & \textrm{~if~} p_e<p_d \\ \infty & \textrm{~if~}  p_e\geq p_d
\end{cases}. 
\end{align*}
\end{lemma}
\noindent Lemma~\ref{l:simp} is proven in Appendix~\ref{alemma}. 
The following corollary characterizes how sensitivity depends on association and gives formulas for $l_\phi$ and $u_\phi$.

\begin{corollary}[Sensitivity conditional on $(p_e,p_d)$]
\label{strength}
Given fixed $(p_e,p_d)$ with $p_e,p_d\in(0,1)$ we consider realizable triples $(p_e,p_d,\alpha)$ for $\alpha\in\{RD,RR,OR\}$. For the purpose of causal inference, 
\setlist[description]{font=\normalfont\itshape}
\begin{description}
\item[(i) Association is necessary] 
If $RD\to0$, $RR\to1$, or $OR\to1$ then $T\to 1$.
\item[(ii) More association is better] the threshold $T$ is decreasing in each of $|RD|$, $|RR-1|$, and $|OR-1|$.
\item[(iii) Sufficient association may exist] If $RD\uparrow u_{RD}$, $RR\uparrow u_{RR}$, or $OR\uparrow u_{OR}$ then \[T\downarrow 1-u_{\phi}, 
\qquad \textrm{where~} 
u_\phi := \sqrt{\min\left\{\frac{p_e(1-p_d)}{p_d(1-p_e)},\frac{p_d(1-p_e)}{p_e(1-p_d)}\right\}}.\]
If $RD\downarrow l_{RD}$, $RR\downarrow l_{RR}$, or $OR\downarrow l_{OR}$ then \[T\downarrow 1+l_\phi 
\qquad \textrm{where~} l_\phi := -\sqrt{\min\left\{\frac{p_ep_d}{(1-p_e)(1-p_d)},\frac{(1-p_e)(1-p_d)}{p_ep_d}\right\}}.\]
\end{description}
\end{corollary}

\noindent Corollary~\ref{strength} is proven in Appendix~\ref{a3}. Figure~\ref{TalphaC} shows a graph of $T$ as a function of $RD$, $RR$, and $OR$ and illustrates Corollary~\ref{strength} in the special case of $p_e=p_d=0.5$. In Corollary~\ref{strength} (iii), 
$u_\phi=1\iff p_e=p_d$, and 
$l_\phi=-1\iff p_e+p_d=1$. Sufficient (positive) association exists when $p_e=p_d$ and sufficient (negative) association exists when $p_e+p_d=1$. Sufficient association conditional on $(p_e,p_d)$ means given any $1-R_p R_r>0$ there exists a level of association sufficient to warrant causal inference. 

\begin{rem} 
Corollary~\ref{strength} (iii) implies the existence of study designs parameterized by $(p_e,p_d,R^2_p,R^2_r)$, with $p_e\neq p_d$, $p_e+p_d\neq 1$, and $R_pR_r>\max\{-l_\phi,u_\phi\}$, so that $\eta=1-R_pR_r\leq T$ for every realizable association $\alpha$.
\end{rem} 

\begin{figure}
\includegraphics[width=3.2in]{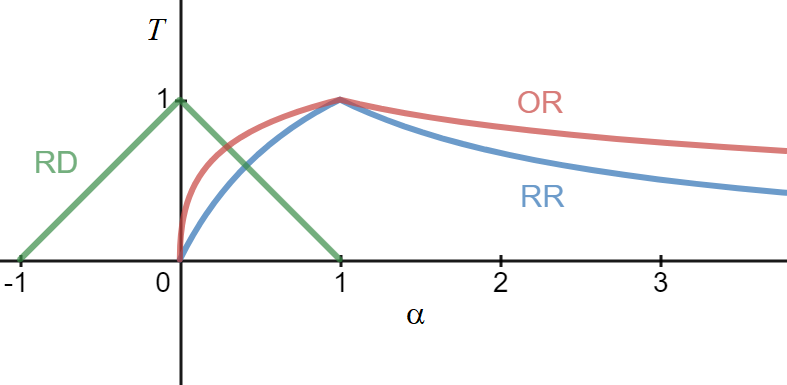} 
\caption{A graph showing $T$ as a function of $\alpha\in\{RD,RR,OR\}$ given $p_e=0.5$ and $p_d=0.5$. $T\to 0$ as $RR\to\infty$ and as $OR\to\infty$.}
\label{TalphaC}
\end{figure}

We say that a study has a \emph{balanced design} if $p_e$ and $p_d$ are near $0.5$. 
One way to measure the \emph{balance} of a study design is to use the quantity 
\begin{equation}
\label{balance}
b=\min\{p_e,(1-p_e),p_d,(1-p_d)\}.
\end{equation}
We can reparametrize $(p_e,p_d)\in (0,1)^2  \setminus \{ (0.5,0.5) \}$ with $(b,\theta)$, where $b\in(0,0.5)$ and $\theta\in[0,2\pi)$ is an angle of rotation. The following corollary characterizes how sensitivity depends on $(p_e,p_d)$ given $OR$.

\begin{corollary}[Sensitivity conditional on $OR$]
\label{cor2}
For the purpose of causal inference, given fixed $OR$ in $(0,\infty)\setminus \{1\}$, and fixed $\theta\in [0,2\pi)$,
\setlist[description]{font=\normalfont\itshape}
\begin{description}
\item[(i) Balance is necessary] If $b\to 0$ then $T\to 1$. 

\item[(ii) More balance is better] The threshold $T$ is decreasing in $b$.

\item[(iii) Sufficient balance does not exist] For any fixed $OR\in(0,\infty)$, \[\displaystyle 0<\inf_{b\in\{(0,0.5)\}}T.\] 
\end{description}
\end{corollary}

\noindent Corollary~\ref{cor2} is proven in Appendix~\ref{a4}. 
The following corollary characterizes how sensitivity depends on $(p_e,p_d)$ given $RD$ and $RR$.

\begin{corollary}
[Sensitivity conditional on $RD$ and $RR$]
\label{cor3}
For the purpose of causal inference 
\setlist[description]{font=\normalfont\itshape}
\begin{description}
\item[(i) Sensitivity conditional on $RD$] For any $RD\in(-1,1)\setminus \{ 0 \}$, on the domain of realizable $(p_e,p_d)$, given fixed $p_e$ the threshold $T$ is decreasing in $|p_d-0.5|$, and given fixed $p_d$ the threshold $T$ is increasing in $|p_e-0.5|$.
\item[(ii) Sensitivity conditional on $RR$]
For any $RR\in(0,\infty)\setminus \{1\}$, on the domain of realizable $(p_e,p_d)$, given fixed $p_e$ the threshold $T$ is decreasing in $p_d$. 
\end{description}
\end{corollary}

\noindent Corollary~\ref{cor3} is proven in Appendix~\ref{a5}. Conditional on $RR$, the dependency of $T$ on $p_e$ given fixed $p_e$ is more complicated.  We demonstrate Corollary~\ref{cor2} (ii) with example applications in Section~\ref{applications}.

\begin{figure}[p]
        \centering
        \begin{subfigure}[b]{0.47\textwidth}
            \centering
            \includegraphics[width=\textwidth]{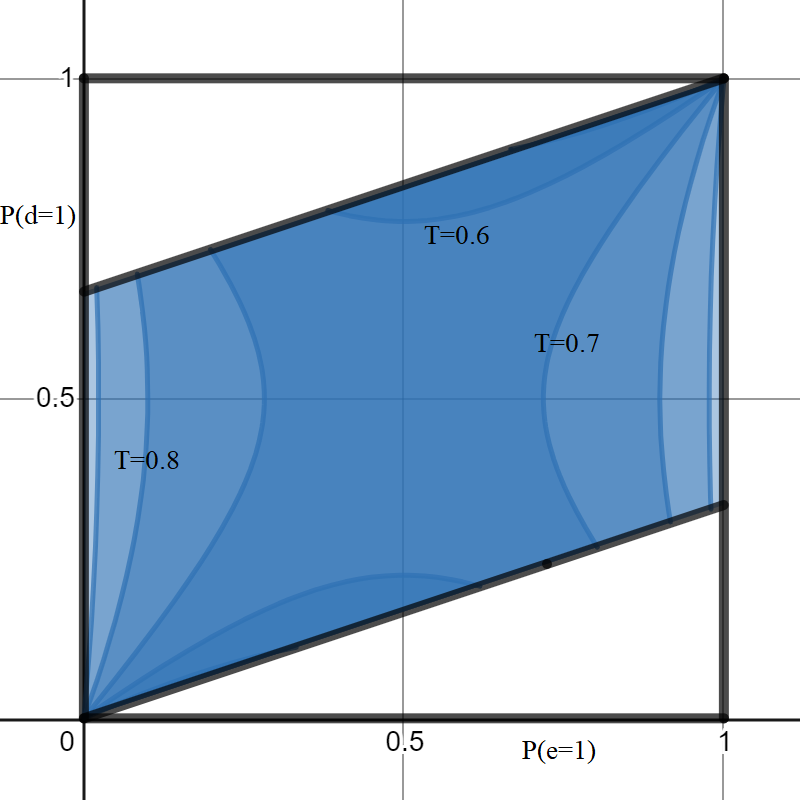}
            \caption[Network2]%
            {{\small $RD=1/3$}}    
            \label{fig:mean and std of net14}
        \end{subfigure}
        \hfill
        \begin{subfigure}[b]{0.47\textwidth}  
            \centering 
            \includegraphics[width=\textwidth]{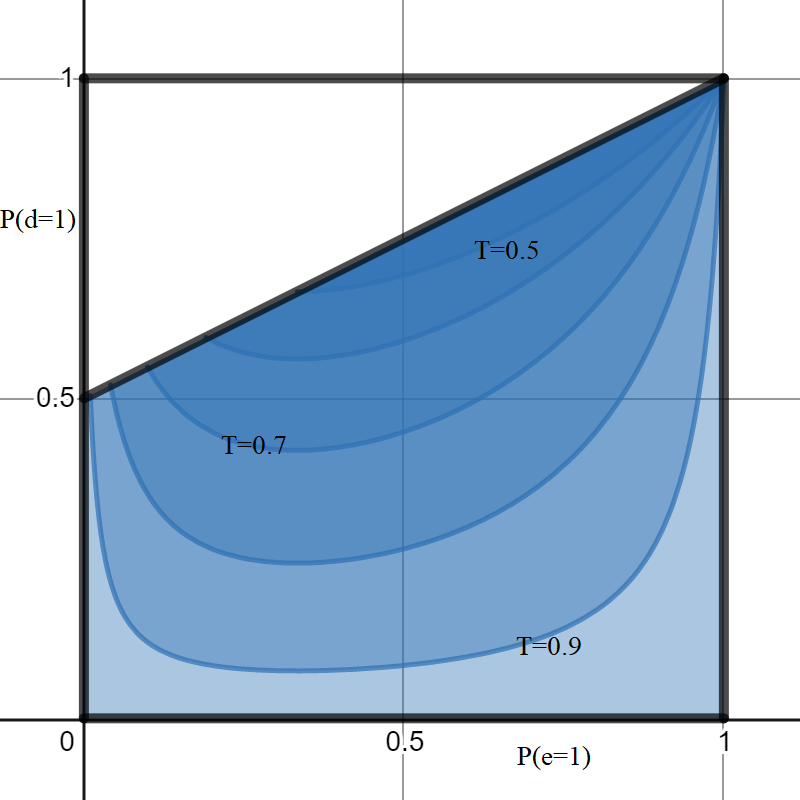}
            \caption[]%
            {{\small $RR=2$}}    
            \label{fig:mean and std of net24}
        \end{subfigure}
        \vskip\baselineskip
        \begin{subfigure}[b]{0.47\textwidth}   
            \centering 
            \includegraphics[width=\textwidth]{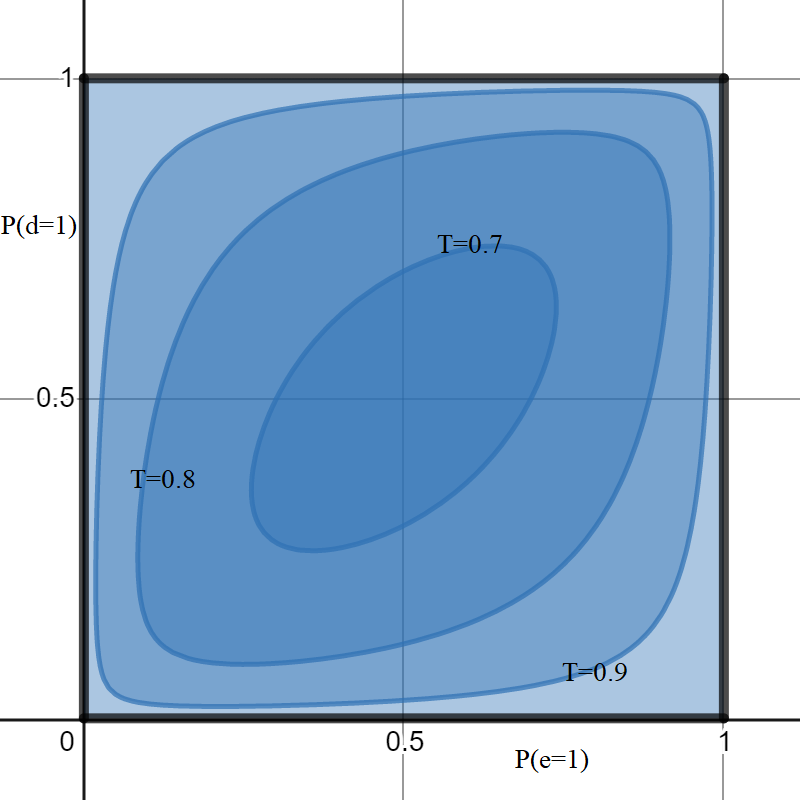}
            \caption[]%
            {{\small $OR=4$}}    
            \label{fig:mean and std of net34}
        \end{subfigure}
        \hfill
        \begin{subfigure}[b]{0.47\textwidth}   
            \centering 
            \includegraphics[width=\textwidth]{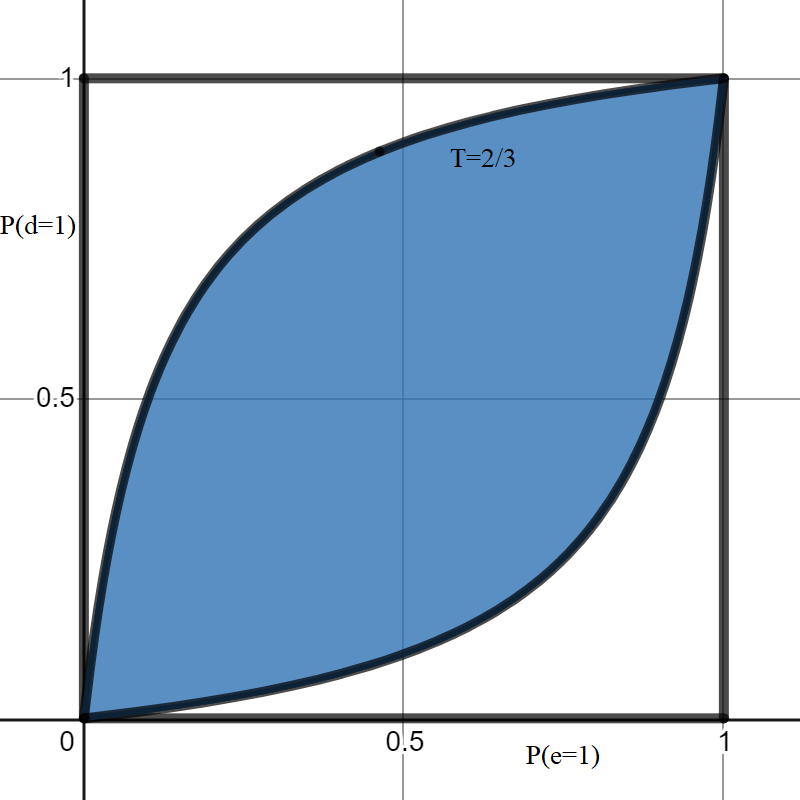}
            \caption[]%
            {{\small $\phi=1/3$}}    
            \label{fig:mean and std of net44}
        \end{subfigure}
        \caption{Conditional on $RD$, $RR$, $OR$, and $\phi$ we plot selected level sets of the sensitivity, $T$, as a function of $(p_e,p_d)$. 
        Note that conditional on $\phi$, sensitivity does not depend on realizable $(p_e,p_d)$.
        At $(p_e,p_d)=(0.5,0.5)$ the values $RD=1/3$, $RR=2$, $OR=4$, and $\phi=1/3$ are all consistent with $p_{01}=1/6$, $p_{11}=2/6$, $p_{00}=2/6$, and $p_{10}=1/6$, and therefore at $(p_e,p_d)=(0.5,0.5)$, by Theorem~\ref{abcd}, all $T$ values are equal to $1-|\phi|=2/3$.  Black curves on the interior of the square indicate the boundaries separating realizable points in blue from unrealizable points in white. }
        \label{RRspace}
    \end{figure}

Corollaries~\ref{cor2} and \ref{cor3} are illustrated in Figure~\ref{RRspace}, where we plot sensitivity conditional on $RD$, $RR$, $OR$, and $\phi$. Conditional on $\phi$ the values of $T$ are constant on the domain of realizable $(p_e,p_d)$. Not all $(p_e,p_d)$ are realizable conditional on $RD$, $RR$, or $\phi$. All $(p_e,p_d)$ are realizable conditional on $OR$.

\section{Example applications}
\label{applications}
Here we use the common measure of association $RR$ (relative risk) and Proposition~\ref{P} to compute the sensitivity $T$ and demonstrate how it depends on more than just strength of association. In our first example application, we analyze a large association only to conclude that it can be explained away with a relatively small $R_pR_r$ value (see Section~\ref{Rdefshere}). In our second example application we analyze a much smaller association and find that a slightly larger $R_pR_r$ value is needed to explain it away. This seeming discrepancy is discussed in more detail within Remark~\ref{rem:ExRem} at the conclusion of this section.

\subsection{Does diabetes cause stroke?}
\label{stroke}
As part of a World Health Organization (WHO) project Multinational Monitoring of Trends and Determinants of Cardiovascular Disease, \citet{Steg} studied $119,733$ Swedish women aged 35--74 for eight years. At the start of this cohort study each woman was tested for diabetes mellitus with a glucose test. Any subsequent stroke, as defined by the WHO, was recorded as a primary outcome for each individual under study. Table~\ref{T1} displays the results. There is a strong, statistically significant, positive association (relative risk RR=5.8; 95\% CI 3.7--6.9) between diabetes and stroke. 

\begin{table}[ht!]
\centering
\caption{A contingency table showing an observed association (RR $=5.8$) between diabetes and stroke.}
\label{T1}
\begin{tabular}{rcc}
\toprule
 & \multicolumn{2}{c}{Diabetes}\\
\cmidrule{2-3}
Stroke&No&Yes\\
\hline
Yes&1,823&647\\
No&110,986&6,277\\
\hline
\end{tabular}
\end{table}

The exposure $e=1$ is the presence of diabetes at the start of the study and the outcome $d=1$ is any stroke occurrence during the subsequent eight year monitoring time period. From the observed data we compute $P(e=1)=0.058$ and $P(d=1)=0.021$. 
Using $P(e=1)=0.058$, $P(d=1)=0.021$, and 
$RR=5.8$ within \eqref{RDformula}  we compute the threshold $T$ of sufficient randomness to be $\textrm{T}=0.87$. The same quantity can be computed using Theorem~\ref{abcd} from the relative frequencies of Table~\ref{T1},
\[p_{01}=0.015,\quad p_{11}=0.005,\quad p_{00}=0.927,\quad p_{10}=0.053.\] 
Assuming the accuracy of our propensity risk model, if $(R^2_p,R^2_r)$ (see Section~\ref{Rdefshere}) satisfies $1-R_pR_r>0.87$ then causal inference is warranted.

\subsection{Does smoking cause COPD?}
\label{COPD}
The Rotterdam Study is a prospective population-based cohort study in the Netherlands. Researchers have been studying a population of roughly 15,000 residents, each aged greater than or equal to $45$ years and living in a well-defined suburb. They are investigating the occurrence of chronic diseases and risk factors. \citet{Terz} have observed a moderate association (relative risk $RR=2.8$) between smoking and chronic obstructive pulmonary disease (COPD). Their raw frequency data is shown in the contingency table shown in Table~\ref{T2}.

\begin{table}[ht!]
\centering
\caption{A contingency table showing an observed association (RR $=2.8$) between smoking and COPD.}
\label{T2}
\begin{tabular}{rcc}
\toprule
 & \multicolumn{2}{c}{Smoking}\\
\cmidrule{2-3}
COPD&No&Yes\\
\hline
Yes&318&1,631\\
No&4679&7,538\\
\hline
\end{tabular}
\end{table}

\newpage
Here the exposure $e=1$ is a history of smoking, and the outcome $d=1$ is COPD diagnosis via extensive examination at a specially built research facility. From the observed data we compute $P(e=1)=0.647$ and $P(d=1)=0.138$. 
Using $P(e=1)=0.647$, $P(d=1)=0.138$, and 
$RR=2.8$ within \eqref{RDformula} we compute the threshold $T$ of sufficient randomness to be $\textrm{T}=0.84$. The same quantity can be computed using Theorem~\ref{abcd} from the relative frequencies of Table~\ref{T2},
\[p_{01}=0.02,\quad p_{11}=0.12,\quad p_{00}=0.33,\quad p_{10}=0.53.\] Assuming the accuracy of our propensity risk model, if $(R^2_p,R^2_r)$ (see Section~\ref{Rdefshere}) satisfies $1-R_pR_r>0.84$ then causal inference is warranted.

\begin{rem} \label{rem:ExRem}
In Section~\ref{stroke} the measured association was $RR_{\textrm{stroke}}=5.8$, and the scientific question is whether diabetes causes stroke. In Section~\ref{COPD} the measured association was $RR_{\textrm{COPD}}=2.8$, and the scientific question is whether smoking causes COPD. Based on the magnitudes of association alone it appears that $RR_{\textrm{stroke}}=5.8$ is more conducive to causal interpretation. However, in Section~\ref{stroke} we computed $T_{\textrm{stroke}}=0.87$, and in Section~\ref{COPD} we computed $T_{\textrm{COPD}}=0.84$. Based on the thresholds it appears that $RR=2.8$ is more conducive to causal interpretation. The threshold $T_{\textrm{stroke}}$ is higher despite $RR_{\textrm{stroke}}=5.8$ in part because of the rare outcome $P(\textrm{stroke}=1)=0.021$. The threshold $T_{\textrm{COPD}}$ is lower despite $RR_{\textrm{COPD}}=2.8$ in part because of the more common outcome $P(\textrm{COPD}=1)=0.138$. See Figure~\ref{RRspace} to appreciate how increasing $p_d$ decreases $T$ conditional on $RR$. 
Thus, the relative risk alone is an inadequate measure of sensitivity, especially when compared with the $\phi$ coefficient of Theorem~\ref{abcd} and Figure~\ref{RRspace}. %
\end{rem}

\section{Discussion}
\label{discussion}
In this paper we have explained how coefficients of determinism $R^2_p$ and $R^2_r$ determine the randomness $\eta$ of the data generating process (see Section~\ref{Rdefshere}). We have also defined a threshold, $T$, of sufficient randomness (see \eqref{SR}) and demonstrated how to compute $T$ from observed data (see Section~\ref{results}). If our propensity-risk model is accurate then $1-R_pR_r>T$ warrants causal inference. 
However, for causal inference it is not necessary to demonstrate $1-R_pR_r>T$, as there are many classic approaches to causal inference with demonstrated utility \citep{Pearce}. Our introduced methodology is complementary and designed to quantify uncertainty in causal inference from natural experiments and quasi-experiments. The example applications of Section~\ref{applications} demonstrate the utility of $T$ as an informative sensitivity parameter. We saw how there can be a lower threshold for causal inference despite smaller measured association; see  Remark~\ref{rem:ExRem}.

In the example applications of Section~\ref{applications} we computed thresholds of sufficient randomness, but we didn't attempt to specify sets of plausible values for $R^2_p$ and $R^2_r$ (see Section~\ref{Rdefshere}). If our propensity-risk model is appropriate, then $1-R_pR_r>T$ is sufficient for causal inference, but there may be uncertainty about the set of plausible $(R^2_p,R^2_r)$ values. Specification of plausible values can be based on knowledge of the processes assigning $e$ and $d$, or knowledge of attributes of individuals in the population under study. 
Care is required when specifying a plausible set for $R^2_r$ because $r$ is risk in the counterfactual absence of exposure (see Section~\ref{methods}). Separate upper bounds for $R_p$ and $R_r$ can be multiplied for comparison with $R^2_\star$ (see \eqref{SR}). If the resulting product is less than $R^2_\star$ then causal inference is warranted. This approach is sharp in the following sense: if there is a distribution $\mu\in\mathcal{P}_0(S)$ with $R^2(\mu)=x$, then for any $(R_p,R_r)$ satisfying $0<R_p,R_r<1$ and $R_p R_r=x$, there exists a distribution $\tilde \mu\in\mathcal{P}_0(S)$ satisfying  
$ \frac{\sigma^2_p}{E[p](1-E[p])}=R^2_p$ and $\frac{\sigma^2_r}{E[r] (1-E[r])}=R^2_r$. 
Thus, information is not lost by using $R^2=R_pR_r$ in place of $(R_p,R_r)$, and this justifies our use of the geometric mean in definitions for $\eta$ (see Section~\ref{Rdefshere}) and $T$ (see Section~\ref{thresh}).

To improve intuition note that increasing variances in propensity or risk can reduce the randomness (see Figure~\ref{sampledistributions}). It is best for causal inference when both distributions of propensities and risks have small variance. It is ideal when there is not variance in propensity nor risk. Instead of asking how exposed and unexposed individuals differ with regards to measured and unmeasured covariate data, we are asking how stochastic (\ie, non-deterministic) are the processes assigning exposure and outcome. A partial answer can be obtained from the field of genetics, where measures of heritability are analogous to coefficients of determination \citep{GC}. Instead of adjusting for conveniently measured covariates we can adjust for unmeasured genomic variables by reviewing the literature to bound measures of heritability for the exposure and disease. 
Note that specification of plausible sets for $R^2_{p}$ and $R^2_r$ is an inexact science, and genomic variables alone are not sufficient for that task, although they can provide lower bounds, and often that will be enough to cast doubt on direct causal inference from the observed association. 
When both the exposure and disease are highly heritable then causal interpretation of an observed association between exposure and disease may be complicated (\cf \citet{PearlDM}).

The following descriptions of studies are examples to help distinguish between studies of insufficient and sufficient randomness. \citet{Almond10} studied the effectiveness of intensive treatment of low-birth-weight newborns, using a regression discontinuity design. They argue that treatment assignment was ``as good as random'', but if measurement error is present then our propensity-risk model indicates that their quasi-experimental design and others like it still have insufficient actual randomness. Perhaps an outcome is more stochastic when longer amounts of time have passed since exposure. While \citet{Almond10} studied infant morality at one week, one month, and one year, we may think of the outcome at one year as being more stochastic, and therefore the association between treatment and that outcome at one year is more conducive to causal inference, all other things equal. Likewise, in studies of the relative age effect or birthdate effect \citep{mpage}, to the extent that success in sports or life requires luck, or chance, the outcome decades later could be highly stochastic, possibly facilitating causal inference. Finally, with randomized trials that allow patients to switch from control to treatment for ethical reasons \citep{Lat}, it is possible to demonstrate sufficient randomness as long as the proportion of patients who switch treatments is not too large. Simply put, the randomized trial need not be perfect---a warrant for causal inference is provided by sufficient randomness.

The optimization problem of \eqref{e:OptProblem} can be reformulated in a variety of equivalent ways. The constraints may be rewritten as $(E[p],E[r],\sigma_{pr}) = \left(p_e,p_r,\sigma_{ed} \right)$,
where $\sigma_{pr}=E[(p-E[p])(r-E[r])]$ is latent covariance and $\sigma_{ed}=p_{01}p_{10}-p_{00}p_{11}$ 
is the observed covariance. We obtain the same optimal solution with 
$|\sigma_{pr}|>|\sigma_{ed}|$ 
in place of $\sigma_{pr}=\sigma_{ed}$. Alternatively, for $\alpha\in\{RD,RR,OR\}$ we may find $\sigma^\star_{pr}=\max\{\sigma_{pr} \colon E[p]=p_e, E[r]=p_d\}$ and then use the monotonicity of spurious $\alpha$ as a function of $\sigma_{pr}$ to compute a maximum spurious $\alpha^\star$ for comparison with observed  $\alpha$. 
Also, we may define $\rho_{ed}=\phi$ which is consistent with $\phi=\sigma_{ed}/\sqrt{p_e(1-p_e)r_e(1-r_e)}$. Then, given $(E[p],E[r],\sigma_{pr})=(p_e,p_d,\sigma_{ed})$ we can factor 
\begin{align}
\nonumber\sigma_{ed}=\sigma_{pr}=\sigma_{p}\sigma_r\rho_{pr}&=R_p\sqrt{E[p](1-E[p])}R_r\sqrt{E[r](1-E[r])}\rho_{pr}\\
&=R_p\sqrt{p_e(1-p_e)}R_r\sqrt{p_d(1-p_d)}\rho_{pr}
\label{linetwo}
\end{align} 
and solve \eqref{linetwo} for $R^2=R_pR_r$ to obtain \begin{equation}
\label{interesting1}
R^2=\phi/\rho_{pr}=\rho_{ed}/\rho_{pr}
\end{equation}
under $H_0$, c.f \eqref{randomness2}. 
From \eqref{interesting1} we then obtain
\begin{equation}
R^2_\star=\rho_{ed}/\max_{\mathcal{P}_0(S)}\{\rho_{pr}\},
\end{equation} 
\cf \eqref{SR}. Finally, we may express $R^2_\star$ using the $\chi^2$ statistic of the chi-squared test of independence as applied to a $2\times 2$ contingency table without a continuity correction. With a finite sample size $n$ we have
{\small
\begin{align} 
    \nonumber \chi^2 &=n\left(\frac{(p_{01}-(1-p_0)r_0)^2}{(1-p_0)r_0}+\frac{(p_{11}-p_0r_0)^2}{p_0r_0}+\frac{(p_{00}-(1-p_0)(1-r_0))^2}{(1-p_0)(1-r_0)}+\frac{(p_{10}-p_0(1-r_0))^2}{p_0(1-r_0)}\right) \\
&=\frac{n(p_{01}p_{10}-p_{11}p_{00})^2}{(p_{01}+p_{00})(p_{11}+p_{10})(p_{01}+p_{11})(p_{00}+p_{10})}.
\label{chiform}
\end{align}}
Asymptotically, by \eqref{SR}, Theorem~\ref{abcd}, and \eqref{chiform}, we have 
\begin{equation}
\label{note}R^2_\star=|\phi|=\sqrt{\chi^2/n}.
\end{equation}
By Theorem~\ref{abcd} and \eqref{note} we see that both the measure of association $\phi$ and the statistic $\chi^2$ inform us of the threshold $T$ independent of $(p_e,p_d)$.

\subsection{Future directions}
\label{fd}
In this paper, we have introduced a notion of randomness in a simplified setting and there are several interesting future directions that further extend the applicability. Under $H_0$ our propensity-risk model assigns $(e,d)$ to individuals, assuming independent assignment. If we eliminate the independence assumption then from any observed $(e,d)$ we can find a feasible latent distribution $\mu\in\mathcal{P}_0(S)$ with \emph{technically} sufficient randomness $1-R^2(\mu)\approx 1$, but \emph{realistically} there would be insufficient randomization if exposure was assigned to the exposed subpopulation at once in mass, for instance. \citet{Fisher} speaks of Randomization as the ``reasoned basis'' for causal inference, and \citet[p. 37]{Rose10} explains how Fisher's phrase is in reference to randomization distributions. We could extend our definition of the randomness into the realm of information theory \citep{Cover} and consider \emph{entropy} as a sensitivity parameter. Causal inference is hindered by lack of independence of observations, insufficient sample size, insufficient randomness of the data generating process, and a lack of balance as defined in \eqref{balance}, and all of these constructs are related to entropy. Our introduced methodology can be adapted for application with smaller sample sizes, and if the sample size is small to moderate we recommend parametric bootstrapping with a multinomial distribution, using the observed probabilities $(p_{01},p_{11},p_{00},p_{10})$ as the parameters, resulting in a bootstrapped distribution of $T$ values. 

This paper has emphasized the stochasticity of the data generating process and de-emphasized conditioning on covariates. One way to incorporate measured covariate data into the analysis would be to formulate the constraints conditional on the covariate levels and express $\mu$ as a mixture distribution. With smaller sample sizes we may utilize a transported algorithm to model $r$ (\cf \citet[Section 2.3]{Rose2002}. The randomness $\eta=1-R^2$ of the data generating process remains the same, and we could still seek to minimize marginal $R^2(\mu)$, thus generalizing the threshold $T$. In some situations conditioning on covariates can increase bias, and it is not always clear how best to proceed \citep{DM}. Our propensity-risk model provides some insight, and a new perspective---conditioning on covariate data is useful to the extent that it lowers the threshold, $T$, of sufficient randomness for causal inference. If our propensity-risk model of the data generating process is accurate, then any conditioning on measured covariate data can only lower $T$, because $R^2_\star$ can only increase if the  constraint set, $\mathcal P_0 (S)$, is further constrained.

\clearpage
\bibliographystyle{abbrvnat}
\bibliography{refs.bib}

\clearpage
\appendix
\section{Mathematical proofs}
\label{s:Proofs}

\subsection{Proof of Theorem~\ref{abcd}}
\label{a1}
Write $\mathcal{P}(S)$ for the space of distributions on $S= (0,1)^2$ and 
$\mathcal P_0 (S) \subset \mathcal P (S)$ for the class of feasible distributions satisfying the constraints in \eqref{e:OptProblemb}--\eqref{e:OptProbleme}. 
For a feasible distribution $\mu \in \mathcal P_0(S)$, the latent covariance is given by 
\begin{subequations}
\label{e:sigmaComp}
\begin{align}
\sigma_{pr} &= \int_S 
\left(p - E[p]\right) \left(r - E[r]\right) \ d\mu \\
&= \int_S pr \ d\mu - E[p] E[r] \\
\label{e:sigmaCompC}
&= p_{11} - \left( p_{10} + p_{11} \right) \left( p_{01} + p_{11} \right) \\
&= p_{11}( 1 - p_{11} - p_{01} - p_{10} ) - p_{10} p_{01} \\
\label{e:sigmaCompE}
&=  p_{11}p_{00} - p_{10} p_{01} \\
\label{sigmas}& =: \sigma_{ed}. 
\end{align}
\end{subequations}
Note that in \eqref{e:sigmaCompC} we used the constraints in \eqref{e:OptProblem} and in \eqref{e:sigmaCompE} we used the fact that $ p_{01} + p_{11} + p_{10} + p_{00}  = 1$.
By the Cauchy-Schwarz inequality and \eqref{e:sigmaComp}, any  $\mu \in \mathcal{P}_0(S)$ satisfies 
$$
\sigma_p^2(\mu) \sigma_r^2(\mu) \geq \sigma_{pr}^2(\mu) = 
\sigma_{ed}^2.
$$
By \eqref{e:OptProblemb}--\eqref{e:OptProbleme}, any $\mu \in P_0(S)$ satisfies 
$$
E[p]  \left(1-E[p]\right)  E[r]  \left( 1-E[r] \right)  
\ =  \ 
p_e(1-p_e)p_d(1-p_d),
$$
where $p_e=P(e=1)$ and $p_d=P(d=1)$.
Thus, we have that any $\mu \in P_0(S)$ satisfies 
\begin{equation} 
\label{e:lowerBound}
R^2(\mu) \geq
\frac{|\sigma_{ed}|}{\sqrt{p_e p_d (1-p_e) (1-p_d)}}.
\end{equation}

It remains to show that there exists a feasible probability distribution, $\mu \in \mathcal P_0(S)$ that  attains the lower bound in \eqref{e:lowerBound}.
Write the signum function
$\sgn(x)=\begin{cases}1 &\textrm{if~} x>0\\ 0 &\textrm{if~} x=0\\ -1&\textrm{if~} x<0\end{cases}$ 
and define 
\begin{align*}
\theta_1 &=\min \left\{p_e,\frac{1}{2}+\sgn \left(\sigma_{ed} \right)\left(p_d-\frac{1}{2}\right)\right\}
=\begin{cases}\min\{p_e,p_d\}&\textrm{~if~}\sigma_{ed}> 0\\\min\{p_e,\frac{1}{2}\}&\textrm{~if~}\sigma_{ed}=0 \\ \min\{p_e,1-p_d\}&\textrm{~if~}\sigma_{ed}<0 \end{cases} \\
\theta_2&=\min \left\{1-p_e,\frac{1}{2}+\sgn \left(\sigma_{ed} \right)\left(\frac{1}{2}-p_d\right)\right\}=\begin{cases}\min\{1-p_e,1-p_d\}&\textrm{~if~}\sigma_{ed}> 0\\\min\{1-p_e,\frac{1}{2}\}&\textrm{~if~}\sigma_{ed}=0 \\ \min\{1-p_e,p_d\}&\textrm{~if~}\sigma_{ed}<0 \end{cases}.
\end{align*}
We claim that the lower bound in \eqref{e:lowerBound} is attained by the two point-mass distribution
\begin{equation*}
\mu^\star(p,r)= \frac{\theta_1}{\theta_1+\theta_2} \delta(p-p_1, r-r_1) + 
\frac{\theta_2}{\theta_1+\theta_2} \delta(p-p_2, r-r_2),
\end{equation*}
where $\delta(p,r)$ is the Dirac delta distribution and 
\begin{align*}
(p_1,r_1) &:=\left(p_e+ k_1, \ p_d+\sgn(\sigma_{ed})
k_1 \right),   \qquad 
k_1 := \sqrt{\frac{\theta_2}{\theta_1}}\sqrt{|\sigma_{ed}|} \\
(p_2,r_2) &:= \left( p_e- k_2, \ p_d-\sgn(\sigma_{ed}) k_2 \right), \qquad 
k_2 := \sqrt{\frac{\theta_1}{\theta_2}}\sqrt{|\sigma_{ed}|}.
\end{align*}

We first claim that $(p_1,r_1)\in S$ and $(p_2,r_2) \in S$ so that $\int_S d\mu^\star = 1$, \ie, $\mu^\star \in \mathcal P(S)$. 
If $\sigma_{ed} = 0$,
then $k_1 = k_2 = 0$, so $(p_1,r_1) = (p_2,r_2) = (p_e,p_d) \in S$.  
Now assume $\sigma_{ed} \neq 0$.
We claim that 
\begin{equation}
    \label{sigbound}
-\min\{p_ep_d,(1-p_e)(1-p_d)\}<\sigma_{ed}<\min\{p_e(1-p_d),p_d (1-p_e)\}.
\end{equation}
To see 
\eqref{sigbound}, we first compute
$\sigma_{pr} 
= \int_S \left(p - p_e \right) \left( r - p_d \right) \ d\mu 
= \int_S p r \ d\mu - p_e p_d$. 
Since $(p,r)\in (0,1)^2$, we have that
$ \int_S p r \ d\mu < \min\{ p_e, p_d \}$, from which the claimed upper bound in \eqref{sigbound} follows. 
We also have that $\int_S pr \ d\mu >0$, so that $\sigma_{pr} > - p_e p_d$. 
Finally from the identity $\sigma_{pr} = \int_S(1-p)(1-r) \ d\mu - (1-p_e)(1-p_d)$ and the observation that  $\int_S(1-p)(1-r) \ d\mu >0$, we have that $\sigma_{pr} > - (1-p_e)(1-p_d)$. Putting these two inequalities together gives the claimed lower bound in \eqref{sigbound}. With $\sigma_{ed}\neq 0$ we have 
\[\theta_1\theta_2=\begin{cases}\min\{p_e(1-p_d),p_d(1-p_e)\}\textrm{~if~}\sigma_{ed}>0\\ \min\{p_ep_d,(1-p_e)(1-p_d)\}\textrm{~if~}\sigma_{ed}<0\end{cases}\]
and from \eqref{sigbound} we then have $\theta_1\theta_2>|\sigma_{ed}|$
so that 
$k_1 = \theta_2  \sqrt{\frac{|\sigma_{ed}|}{\theta_1 \theta_2}} 
< \theta_2$ and 
$k_2 = \theta_1  \sqrt{\frac{|\sigma_{ed}|}{\theta_1 \theta_2}}
< \theta_1$.
Because $k_1<\theta_2$ and $k_2<\theta_1$ we have $(p_1,r_1)\in S$ and $(p_2,r_2) \in S$. 

We now verify that $\mu^\star$ satisfies the constraints in \eqref{e:OptProblemb}--\eqref{e:OptProbleme}. We first compute
$$
\int_S pr \ d\mu^\star = \frac{\theta_1}{\theta_1+\theta_2} p_1 r_1 + \frac{\theta_2}{\theta_1+\theta_2} p_2 r_2 = p_e p_d + \sigma_{ed} = p_{11},
$$
which verifies \eqref{e:OptProblemc}. We then compute  
$$
\int_S p \ d\mu^\star = 
\frac{\theta_1}{\theta_1+\theta_2} p_1 + \frac{\theta_2}{\theta_1+\theta_2} p_2
= p_e + \frac{\theta_1 k_1 - \theta_2 k_2}{\theta_1 + \theta_2} 
= p_e 
\quad
\textrm{and similarly} 
\quad
\int_S r \ d\mu^\star = p_d.
$$
This gives us that 
\begin{align*}
& \int_S (1-p)r \ d\mu^\star = p_d - p_{11} = p_{01} \\
& \int_S p (1-r) \ d\mu^\star = p_e - p_{11} = p_{10} \\
& \int_S (1-p)(1-r) \ d\mu^\star = 1 - p_e - p_d + p_{11} = p_{00},
\end{align*}
which verifies \eqref{e:OptProblemb}, \eqref{e:OptProblemd}, and \eqref{e:OptProbleme}.

To evaluate $R^2(\mu^\star)$, we compute
$$
\sigma_p^2 = \frac{\theta_1}{\theta_1+\theta_2} (p_1 - p_e)^2 + 
\frac{\theta_2}{\theta_1+\theta_2}
(p_2 - p_e)^2 
= \frac{\theta_1 k_1^2 + \theta_2 k_2^2}{\theta_1+\theta_2}
= | \sigma_{ed} | 
$$
and similarly $\sigma_r^2= | \sigma_{ed} | $. Then 
$$
R^2(\mu^\star) = \frac{ |\sigma_{ed} | }{\sqrt{ p_e  \left(1- p_e \right)  
p_d  \left( 1- p_d\right)  
}}
$$
which together with \eqref{e:lowerBound} establishes that $\min_{\mu \in \mathcal P_0(S) } R^2(\mu) = |\phi|$. From the definition of $T$ in \eqref{SR}, this establishes \eqref{tform}.
\hfill $\blacksquare$

\subsection{Proof of Proposition~\ref{P}}
\label{a2}
For each measure of association, $\alpha \in \{RD,RR,OR\}$, we derive the formula in \eqref{RDformula}. 

\medskip

\noindent \underline{$\alpha = RD$.}
By definition, 
$RD=\frac{p_{11}}{p_{11}+p_{10}}-\frac{p_{01}}{p_{00}+p_{01}}$ so 
\begin{align*}
p_e(1-p_e)RD
= (p_{00}+p_{01}) p_{11} - (p_{11}+p_{10})   p_{01} 
= p_{00}p_{11} - p_{10} p_{01} 
= \phi \sqrt{p_e (1-p_e) p_d (1-p_d)}
\end{align*}
So we have that $k |RD|=  |\phi| $, so $T = 1 - |\phi| = 1 - |RD| k$, as desired.

\bigskip

\noindent \underline{$\alpha = RR$.} 
Using the definition of $RR$, we have 
\begin{align*}
RR - 1 &= \frac{p_{11}}{p_e} - \frac{1-p_e}{p_{01}} - 1 
= 
\frac{p_{11} - p_e p_d}{p_e p_{01}} \\ 
p_d(RR-1) &= \frac{p_d}{p_e}\frac{p_{11} - p_e p_d}{p_{01}}\\
1 + p_e (RR-1) & = \frac{p_d(1-p_e)}{p_{01}}.
\end{align*}
Dividing these last two quantities, we obtain
$$
\frac{p_d(RR-1)}{1 + p_e (RR-1)} = 
\frac{p_{11} - p_e p_d}{p_e(1-p_e)}. 
$$
On the other hand, 
$$RD = \frac{p_{11}}{p_e} - \frac{p_{01}}{1-p_e} = \frac{p_{11}(1-p_e) - p_{01}p_e}{p_e(1-p_e)} 
= \frac{p_{11} - p_e p_d}{p_e(1-p_e)}.
$$
Combining these last two expressions, we have that 
\begin{equation*}
RD=\frac{ p_d(RR-1)}{1+p_e(RR-1)},
\end{equation*} 
which together with the expression  $T = 1 - |RD|k$ gives the desired result.

\bigskip

\noindent \underline{$\alpha = OR$.} 
We compute 
\begin{align}
\label{OR/RR}
    \frac{OR}{RR} 
       & = \frac{p_{00}}{p_{10}}\frac{(p_{10}+p_{11})}{(p_{01}+p_{00})}\\
       \nonumber
       & = \frac{p_{10}p_{00}+ p_{10}p_{01} - p_{10}p_{01}+p_{00}p_{11}}{p_{10}p_{01}+p_{10}p_{00}}\\
       \nonumber
       & = 1-\frac{p_{10}p_{01}}{p_{10}p_{00}+p_{10}p_{01}} + \frac{p_{00}}{p_{10}} \left(\frac{p_{11}}{p_{00}+p_{01}} \right)\\ 
       \nonumber
       & = 1-\frac{p_{01}}{p_{00}+p_{01}} + \frac{p_{01}}{p_{00} + p_{01}}OR. 
\end{align}
Rearranging, we obtain
\begin{equation}
\label{RROR}
RR=\frac{OR}{1-\frac{p_{01}}{p_{01}+p_{00}}+\frac{p_{01}}{p_{01}+p_{00}}OR}.
\end{equation}
We proceed to express $\frac{p_{01}}{p_{01}+p_{00}}$ in terms of $RR$, $p_e$, and $p_d$. By the definition of $RR$ we have
\begin{equation}\label{4comb}RR=\frac{p_{11}}{p_{01}}\frac{(1-p_e)}{p_e}.\end{equation}
From \eqref{4comb} and $p_d=p_{01}+p_{11}$ we solve for 
\[p_{01}=\frac{p_d(1-p_e)}{1+p_e(RR-1)}\]
and then divide by $p_{01}+p_{00}=1-p_e$ to obtain
\begin{equation}\label{4ac}\frac{p_{01}}{p_{01}+p_{00}}=\frac{p_d}{1+p_e(RR-1)}.\end{equation}
By combining \eqref{RROR} and \eqref{4ac} we obtain
\begin{equation}
\label{quadrat}
p_e RR^2+(p_d(OR-1)+(1-p_e)-p_eOR)RR-(1-p_e)OR=0.
\end{equation}
Writing  $a = p_d\left(OR-1\right)+\left(1-p_e\right)-p_eOR$ and $c = (p_e-1)OR$, we apply the quadratic formula to find that 
$$
RR= u_\pm := \frac{-a \pm \sqrt{a^2-4 p_e c}}{2 p_e}.
$$
The discriminant is clearly positive since 
$a^2-4 p_e c = a^2 + 4 p_e(1-p_e)OR \geq 0$ so we have two real roots. 
To choose the sign, we insist that $RR=1$ when $OR=1$, which requires the plus sign in the quadratic formula. Now combining the result with the formula for $T(\alpha)$ with $\alpha = RR$ in \eqref{RDformula}, we obtain the desired formula.  \hfill $\blacksquare$

\subsection{Proof of Lemma~\ref{l:simp}}
\label{alemma}
We can express the measures of association $RD$, $RR$, and $OR$ in terms of the covariance $\sigma_{ed} := p_{11}p_{00} - p_{10} p_{01}$ as follows:
\begin{subequations}
\label{3Rs}
\begin{align}
 RD &=\frac{p_e p_d + \sigma_{ed}}{ p_e p_d +p_e (1- p_d)}-\frac{(1 - p_e)p_d-\sigma_{ed}}{(1 - p_e)p_d + (1- p_e) (1- p_d)}\\
 RR&=\frac{p_e p_d + \sigma_{ed}}{ p_e p_d +p_e (1- p_d)}\frac{(1 - p_e)p_d + (1- p_e) (1- p_d)}{(1 - p_e)p_d-\sigma_{ed}}\\
 OR&=\frac{(p_e p_d + \sigma_{ed})( (1- p_e) (1- p_d) + \sigma_{ed})}{(p_e (1- p_d) - \sigma_{ed})((1 - p_e)p_d - \sigma_{ed})}.
\end{align}
\end{subequations}
From \eqref{sigbound}, we have $\sigma_{ed} \in (l_\sigma,u_\sigma)$, where
$$
l_\sigma = -\min\{p_ep_d,(1-p_e)(1-p_d)\}
\qquad \textrm{and} \qquad
u_\sigma := \min\{p_e(1-p_d),p_d (1-p_e) \}.
$$
Moreover, it is possible to construct sequences of contingency tables such that $\sigma_{ed}$ approaches these lower and upper bounds.  
Now, using the continuity and monotonicity of each of $RD$, $RR$, and $OR$ in $\sigma_{ed}$, 
we can compute upper and lower bounds for each of these measures of association 
\begin{subequations}
\label{six}
\begin{align}
u_{RD}&=\lim_{\sigma_{ed}\to u_\sigma}RD=
\begin{cases}
&\frac{p_d}{p_e}\textrm{~if~} p_e\geq p_d\\
&\frac{1-p_d}{1-p_e}\textrm{~if~} p_e<p_d
\end{cases},\\
l_{RD}&=\lim_{\sigma_{ed}\to l_\sigma}RD=
\begin{cases}
&\frac{p_d-1}{p_e}\textrm{~if~} p_e+p_d\geq 1\\
&\frac{p_d}{p_e-1}\textrm{~if~} p_e+p_d<1
\end{cases},\\
u_{RR}&=\lim_{\sigma_{ed}\to u_\sigma}RR=\begin{cases}&\infty\textrm{~if~} p_e\geq p_d\\&\frac{1-p_e}{p_d-p_e}\textrm{~if~} p_e<p_d\end{cases},\\
l_{RR}&=\lim_{\sigma_{ed}\to l_\sigma}RR=\begin{cases}
&\frac{p_e+p_d-1}{p_e}\textrm{~if~} p_e+p_d\geq 1\\
&0 \textrm{~if~} p_e+p_d<1
\end{cases},\\
u_{OR}&=\lim_{\sigma_{ed}\to u_\sigma}OR=\infty, \textrm{~and}\\
l_{OR}&=\lim_{\sigma_{ed}\to l_\sigma}OR=0.
\end{align}
\end{subequations}
Combining the cases completes the proof.
\hfill $\blacksquare$

\subsection{Proof of Corollary~\ref{strength}}
\label{a3}
From \eqref{RDformula}, with $k=\sqrt{\frac{p_e(1-p_e)}{p_d(1-p_d)}}$, we have
\begin{equation*}
    T(\alpha)=
    \begin{cases}
 1-|RD|k & \textrm{if } \alpha=RD\\
  1-|\frac{p_d(RR-1)}{1+p_e(RR-1)}|k & \textrm{if } \alpha=RR\\
  1-|\frac{p_d(u-1)}{1+p_e(u-1)}|k & \textrm{if } \alpha=OR\\
\end{cases},
\end{equation*}
where $u=\frac{-a+\sqrt{a^2-4 p_e c}}{2 p_e}$, $a = p_d\left(OR-1\right)+\left(1-p_e\right)-p_eOR$, and $c = (p_e-1)OR$. It can be checked that $u=1$ at $OR=1$.

\begin{description}
\item[(i) Association is necessary] Since $T$ is a continuous function of $RD$, $RR$, and $OR$, we have
\begin{description}
   \item[$\alpha = RD$] \quad  $\lim_{RD\to0}T(RD)=T(0)=1$,
   \item[$\alpha = RR$] \quad  $\lim_{RR\to1}T(RR)=T(1)=1$, 
   \item[$\alpha = OR$] \quad  $\lim_{OR\to1}T(OR)=T(1)=1$. 
\end{description}

\bigskip

\item[(ii) Stronger associations are better]   
For $\alpha = RD$, since $k>0$ it is clear that $T$ is decreasing in $\left| RD\right|.$ 

For $\alpha = RR$, from \eqref{RDformula}, we consider two cases: $RR>1$ and $RR<1$. If $RR>1$, we have 
$$
T(RR) = 1-\frac{p_d |RR-1| }{1+p_e|RR-1|} k = f_1(|RR-1|),
$$ 
where $f_1(x) = 1 - \frac{p_d x }{1+p_e x} k$. It is easy to check that $f_1'(x)<0$ for $x>0$. Similarly, for $RR<1$, we have 
$$
T(RR) = 1- \frac{p_d |RR-1| }{1-p_e|RR-1|} k = f_2(|RR-1|),
$$ 
where $f_2(x) = 1 + \frac{p_d x }{1-p_e x} k$. Again, it is easy to check that $f_2'(x)<0$ for $x>0$. 

For $\alpha =OR$, we use the result for $\alpha=RR$,
that RR and OR are monotonically varying together (see \eqref{OR/RR}), and $RR=1$ if and only if $OR=1$.

\item[(iii) Sufficient association may exist]  
This proof follows from Lemma~\ref{l:simp} and its proof. Recall the definition from \eqref{sigmas} that $\sigma_{ed}:= p_{11}p_{00} - p_{10} p_{01}$. Rewriting \eqref{tform} results in 
\begin{equation}\label{trep}T=1-\frac{|\sigma_{ed}|}{\sqrt{p_ep_d(1-p_e)(1-p_d)}}.\end{equation} If $RD\uparrow u_{RD}$, $RR\uparrow u_{RR}$, or $OR\uparrow u_{OR}$ then by \eqref{3Rs} and \eqref{six} we have 
$\sigma_{ed}\uparrow u_\sigma=\min\{p_e(1-p_d),p_d (1-p_e)\}$, 
and from \eqref{trep} we conclude \[T\downarrow 1-\sqrt{\min\left\{\frac{p_e(1-p_d)}{p_d(1-p_e)},\frac{p_d(1-p_e)}{p_e(1-p_d)}\right\}}=1-u_{\phi},\]
as desired.
Likewise, if $RD\downarrow l_{RD}$, $RR\downarrow l_{RR}$, or $OR\downarrow l_{OR}$ then by \eqref{3Rs} and \eqref{six} we have $\sigma_{ed}\downarrow l_\sigma=-\min\{p_ep_d,(1-p_e)(1-p_d\}$, and from \eqref{trep} we conclude \[T\downarrow 1-\sqrt{\min\left\{\frac{p_ep_d}{(1-p_e)(1-p_d)},\frac{(1-p_e)(1-p_d)}{p_ep_d}\right\}}=1+l_\phi,\]
as desired.
\hfill $\blacksquare$
\end{description}

\subsection{Proof of Corollary~\ref{cor2}}
\label{a4}
In this proof, we will use symmetry properties of $T$ and $OR$ several times. 
$T$ and $OR$ are both invariant under either of the following transformations: 
\begin{align*}
   &( p_{01}, p_{10}) \mapsto ( p_{10},p_{01}) \\
   &  (p_{00}, p_{11}) \mapsto ( p_{11},p_{00}).
\end{align*}
Also, either of the transformations $p_e\mapsto (1-p_e)$ and $p_d\mapsto (1-p_d)$ take $OR$ to $1/OR$.
These can seen from \eqref{tform} and the definition of the odds ratio. 

\begin{description}
\item[(i) Balance is necessary] 
Using the above symmetry properties of $T$ and $OR$, without loss of generality, we may assume that $OR> 1 \iff RR \geq 1$ and $p_d \geq p_e$. Rearrangement of \eqref{quadrat} gives
 \[p_e (RR-1)^2+ \left(1+(p_d-p_e)(OR-1) \right)(RR-1) - (1-p_d)(OR-1)=0. \]
With $h=(OR-1)$ we then have
\[RR-1=\frac{-(1+(p_d-p_e)h)+\sqrt{(1+(p_d-p_e)h)^2+4hp_e(1-p_d)}}{2p_e}.\]
Then from \eqref{RDformula} with $\alpha = RR$ and assuming that $RR\geq 1$ we have
\begin{equation}
\label{gformula}
1-T=\frac{\sqrt{(1+(p_d-p_e)h)^2+4hp_e(1-p_d)}-(1+(p_d-p_e)h)}{2+\sqrt{(1+(p_d-p_e)h)^2+4hp_e(1-p_d)}-(1+(p_d-p_e)h)}\sqrt{\frac{p_d(1-p_e)}{p_e(1-p_d)}}.
\end{equation}
If either $p_e\to 0$ or $p_d\to 1$ then $T\to 1$ because, by l'H\^{o}pital's rule,
\begin{equation*}
\frac{\sqrt{(1+(p_d-p_e)h)^2+4hp_e(1-p_d)}-(1+(p_d-p_e)h)}{\sqrt{p_e(1-p_d)}}\to 0,
\end{equation*}
and
\[\frac{\sqrt{p_d(1-p_e)}}{2+\sqrt{(1+(p_d-p_e)h)^2+4hp_e(1-p_d)}-(1+(p_d-p_e)h)}\]
is bounded.

\item[(ii) More balance is better]
Define $v_1=p_e+p_d-1$ and $v_2=p_e-p_d$ so that 
$p_e=(v_1+v_2+1)/2$ and $p_d=(v_1-v_2+1)/2$. 
Using the above symmetry properties of $T$ and $OR$, without loss of generality, we may assume that $OR> 1 \iff RR > 1$, $v_1\geq0$, $v_2\geq 0$, and $v_1 + v_2 < 1$. It suffices to show, for $i=1,2$, that $1-T$ is strictly decreasing in $v_i$ if $v_i>0$.

Writing $h=OR-1$, we obtain from \eqref{gformula},
\begin{align*}
1-T 
&=\frac{\sqrt{1+h((h+1)v_2^2+1-v_1^2)}+hv_2-1}{\sqrt{1+h((h+1)v_2^2+1-v_1^2)}+hv_2 +1} \sqrt{\frac{ (1 - v_2)^2 -v_1^2}{(1 + v_2)^2 -v_1^2}} \\
&= \frac{A-1}{A +1} B,
\end{align*}
where $A := \sqrt{1+h((h+1)v_2^2+1-v_1^2)}+hv_2>1$ and $B := \sqrt{\frac{ (1 - v_2)^2 -v_1^2}{(1 + v_2)^2 -v_1^2}}>0$.
We compute 
\begin{equation}\label{two} \frac{\partial}{\partial v_i} (1-T) = 
\frac{2B }{ (A+1)^2}  \frac{\partial A}{\partial v_i} + \frac{A-1}{A +1} \frac{\partial B}{\partial v_i}, \qquad i=1,2.
\end{equation}
If $v_1>0$ then
\begin{align*}
\frac{\partial A}{\partial v_1} & = - \frac{hv_1} {A} < 0 \\
\frac{\partial B}{\partial v_1} & = - \frac{v_1(1/B-B)}{(1+v_2)^2-v_1^2} < 0,
\end{align*}
and $1-T$ is strictly decreasing in $v_1$.

We also have
\begin{align*}
\frac{\partial A}{\partial v_2} &=
h + \frac{h(h+1)v_2}{A}>0 \\
\frac{\partial B}{\partial v_2} &= - \frac{(1 - v_2)\frac{1}{B}+(1 + v_2)B}{(1+v_2)^2-v_1^2}
< 0. 
\end{align*}
If $v_2>0$ then we use \eqref{two} and show that 
$f> 0$, where 
\begin{equation*}
f:= - A \frac{(A+1)^2}{2B} \frac{\partial }{\partial v_2}
(1-T) 
= A(A^2-1)C -A h-h(h+1)v_2
\end{equation*}
and
$$
C:=- \frac{1}{2B} \frac{\partial B}{\partial v_2}  = \frac{1-v_1^2-v_2^2}{\left((1-v_2)^2-v_1^2\right)\left( (1+v_2)^2-v_1^2 \right)} \geq 1.
$$
We will first show that $f$ is increasing in $v_2$, \ie, $\frac{\partial f}{\partial v_2} > 0$. 
Using 
$A(A^2-1) \geq 0$ 
and 
$\frac{\partial C}{\partial v_2} \geq 0$, we compute 
\begin{subequations}
\label{e:fv2}
\begin{align}
\frac{\partial f}{\partial v_2} &= 
[ (3 A^2 - 1) C - h ] \frac{\partial A}{\partial v_2}
+ A (A^2 - 1) \frac{\partial C}{\partial v_2} - h (h+1) \\
& \geq [ ( A^2 - 1) C - h ] \frac{\partial A}{\partial v_2}
+ 2 A^2 C \frac{\partial A}{\partial v_2}
 - h (h+1). 
\end{align}
\end{subequations}
We now use the inequality 
$$
A^2-1 = h(v_2^2 + 1 - v_1^2 + 2 v_2 A) \geq h ( 1- v_1^2) 
$$
and the fact that $ C \geq \frac{1}{1-v_1^2}$ to conclude that 
$(A^2 - 1 ) C \geq h$. 
From \eqref{e:fv2}, $\frac{\partial A}{\partial v_2} \geq h$, and $C\geq 1$ we have 
\begin{align*}
\frac{\partial f}{\partial v_2} 
& \geq 2 (h+C) h 
 - h (h+1) \geq h (h+1) > 0. 
\end{align*}
Since $f$ is increasing in $v_2$ and $f \big|_{v_2 = 0} = 0$ we have $f>0$ if $v_2>0$.
\item[(iii) Sufficient balance does not exist]
If $OR=1$ then $T=1>0$ for each $(p_e,p_d)\in(0,1)^2$. If $OR \neq 1$, by Corollary~\ref{cor2}(ii) we have that $T$ is decreasing in $b$ for any $b\in (0,0.5)$. Using the continuity of $T$ in $b$, we have that  \[\inf_{b\in(0,0.5)}T=
T(p_e = 0.5,p_d = 0.5).\]
From \eqref{gformula} we know for any $OR > 1$ that 
\begin{equation*}
1 - T(p_e = 0.5,p_d = 0.5) =\frac{\sqrt{OR}-1}{\sqrt{OR}+1}<1,
\end{equation*} 
which implies that $T(p_e=0.5,p_d=0.5)>0$. 
By symmetry, the result also holds for $OR<1$. 
\hfill $\blacksquare$
\end{description}

\subsection{Proof of Corollary~\ref{cor3}}
\label{a5}
\begin{description}
\item[(i) Sensitivity conditional on $RD$]
We have fixed $RD\in(-1,1)\setminus \{0\}$. From \eqref{RDformula}, we have
\begin{align*}
T = 1 - \left|RD \right| k
= 1 - \left|RD \right|
\sqrt{p_e(1-p_e)} 
\frac{1}{\sqrt{p_d(1-p_d)}}
= 1 - \left|RD \right|
\sqrt{p_e(1-p_e)} 
\frac{1}{\sqrt{ \frac{1}{4} - \left| p_d - \frac{1}{2}\right|^2  }},
\end{align*}
which is decreasing in $\left| p_d - \frac{1}{2}\right|$.
Similarly, we have
\begin{align*}
T = 1 - \left|RD \right|
\sqrt{ \frac{1}{4} - \left| p_e - \frac{1}{2}\right|^2 }
\frac{1}{\sqrt{p_d(1-p_d)}},
\end{align*}
which is increasing in $\left| p_e - \frac{1}{2}\right|$. 

\item[(ii) Sensitivity conditional on $RR$]
We have fixed $RR\in(0,\infty)\setminus \{1\}$. From \eqref{RDformula}, we have 
\begin{align*}
T = 1 - \left|\frac{p_d(RR-1)}{1+p_e(RR-1)} \right| k
= 1 - \left|\frac{(RR-1)}{1+p_e(RR-1)} \right|
\sqrt{p_e(1-p_e)} 
\sqrt{\frac{p_d}{(1-p_d)}},
\end{align*}
which is decreasing in $p_d$ on $(0,1)$.   
\hfill $\blacksquare$
\end{description}

\end{document}